Double domain resonance in a periodically poled ferroelectric wafer.

Igor Ostrovskii and Lucien Cremaldi

The University of Mississippi, Department of Physics and Astronomy,

University, MS 38677, U.S.A.

*Abstract*:

The periodically poled ferroelectric wafer is a two-dimensional phononic superlattice. The important applications of such a solid include novel ultrasonic transducers at the micro/nano-scale for low intensity ultra-sonography, ferroelectric data storage, and development of very high frequency chips for next generation communication and information technologies, and others. Usually, the transformations between the electrical and mechanical energies are required in the applications. In this work, we find the frequency characteristics for an effective acousto-electric interaction in a two-dimensional phononic superlattice inserted in $LiNbO_3$ wafer.

**1**. *Introduction.* The ferroelectric domains/walls are of great interest in science and technology[1-3]. The research on ferroelectric domains resulted in important new fundamental knowledge including structure and energetics of the interdomain walls [4] and their interaction with biased scanning probe microscopy tips [5], and combined elementary excitations of the type of phonon-plasmon-polariton at the interface of a piezoelectric metamaterial and semiconductor [6], to mention a few. The inversely poled ferroelectrics are of great interest in connection with the chips transforming electrical energy into mechanical and vice versa[7], and a very attractive idea of a data recording on ferroelectric[8]. The domain dynamical properties are responsible for efficiency of the acousto-electric transducers, filters and oscillators, inversely poled optical waveguides for laser light manipulation, and poling process itself; all above may well depend on the electrical properties of a superlattice. The latter suggests a problem of electrical current through a multidomain structure. It becomes more important for a practical case of a two-dimensional inversely poled ferroelectric such as wafer or film. That structure is an acoustic superlattice having a so called "stop-band."[9] Here we show that a rf-current flowing through a periodically poled ferroelectric wafer has two maxima versus frequency. The previously known single "domain resonance"[10-12] is valid for a one-dimensional periodically poled ferroelectric only. Mismatching between the operational frequency of any application and resonance frequencies of a superlattice may strongly influence a performance of various chips. Taking into account correct frequencies of the acousto-electric resonances allow for significant progress in development of ultrasonic transducers for medicine and ultrasonics[13-15], high frequency electronics and filters[16] using ferroelectric structures, energy harvesting ferroelectric devices[17], and almost any multidomain application.

**2. *Theory*.** The geometry of the ferroelectric phononic superlattice (FPS) and experimental setup are shown in fig.1. Sample 1 contains N "**a**"-domains with positive piezoelectric coefficient (e > 0) and N "**b**"-domains with negative coefficient e < 0, each domain has length (d). The rf voltage $V(\omega)$ is applied to input electrode 2, and resultant current is detected by electrode 3. To determine the frequency range of the first stopband, the Finite Element Code described in reference[9] was run with the corresponding material constants and parameters of the superlattice. The solution is based on the Hamilton's variational principle, where the total energy functional (1) is minimized.

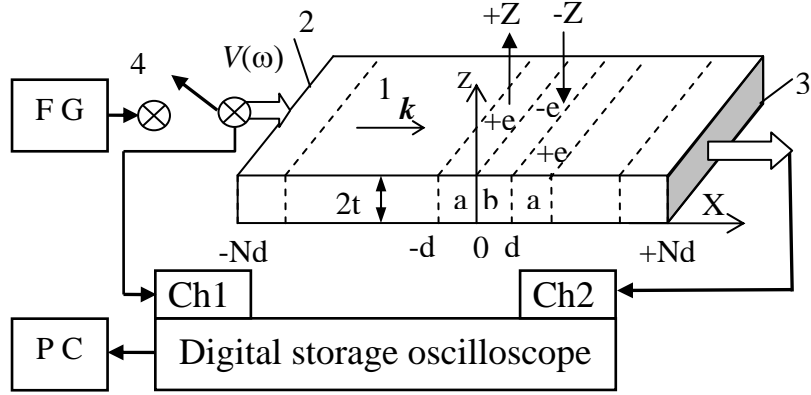

Fig 1. Inversely poled superlattice in ferroelectric wafer, and experimental schematic. 1 – The ZX-cut LiNbO$_3$ wafer with periodically poled domains (a) and (b), 2 – input metal electrode for applying rf voltage $V(\omega)$, 3 – output metal electrode for reading electric current, 4 – coaxial connector,  FG – function generator, PC – computer.

$$\delta\left[\int_C (W_K - W_S + W_D + W)dt + \int_A (W_{D,A})dt\right] = 0 \quad , \quad (1)$$

where, the first integral is taken over crystal bulk, and the second - over the air surrounding sample. The components of total energy per unit width in functional (1) are kinetic energy $W_K$ of an acoustic wave (AW) in the plate, energy of electric field inside crystal $W_D$ and in surrounding air $W_{D,A}$, elastic energy $W_S$ of crystal deformation by AW, and energy W provided by an external source exciting AW; more details are in reference[9]. To compute the functional (1) it is necessary to know all energies in equation (1), which can be found through the equations of motion and electrodynamics. In the configuration of fig.1, the voltage $V(\omega)$ produces an electric field $E_m(\omega)$ inside the wafer that in turn generates AW with the displacement $A_k(\omega)$ and mechanical strain $S_{k\,l} = (dA_k / dx_l)$. Therefore the independent variables are $E_m(\omega)$ and $S_{k\,l}= (dA_k/dx_l)$. There are four known pairs of equations embedding elastic and piezoelectric tensors[18]; and for the case under consideration, the correct choice of the equations of motion (2) and electrodynamics (3) is below:

$$\rho \frac{\partial^2 A_i}{\partial t^2} = \frac{\partial T_{ij}}{\partial x_j} = \frac{\partial}{\partial x_j}[c^E_{ijkl}\frac{\partial A_k}{\partial x_l} - e^{-+}_{mij}(x)E_m] \qquad (2)$$

$$\frac{\partial D_i}{\partial x_i} = \frac{\partial}{\partial x_i}[e^{-+}_{ikl}(x)\frac{\partial A_k}{\partial x_l} + \varepsilon^S_{ij}E_j]=0 \quad , \qquad (3)$$

where, $T_{ij}$ are stress tensor components, $D_i$ is electric displacement, $c^E_{ijkl}$ are elastic modules, $\varepsilon^S_{ij}$ is dielectric permittivity tensor, $e^{-+}_{ikl}(x)$ is piezoelectric constant with negative/positive sign depending on *X*. It is important to note that in different publications on the inversely poled ferroelectrics one can find the dissimilar initial equations (2) and (3). That is why the previously published results, including multidomain ultrasonic transducers/vibrators and inter-domain wall dynamics, cannot be applied for the present problem. The boundary conditions of the zero acoustic stress and zero electric charge at free plate surfaces are applied. The mesh for FEM-computation consists of linear triangle elements with equal sides of length 0.05mm along the *X* and *Z* axes. This mesh is extended into the air above and below the wafer by ten plate thicknesses. The FEM-codes return the frequencies and wave vectors at which the functional (1) is satisfied. The collection of these solutions forms a dispersion curve of propagating zero antisymmetric mode, which is presented in fig.2. The stop-band appears in the frequency range of $F_1$=3.27 MHz to $F_2$=3.67 MHz, where the dispersion curve is interrupted, that range corresponds exactly to AW wavelength $\lambda = 2d = 0.9$ mm, or $(1/\lambda) = 1.11$ mm$^{-1}$. The horizontal line $(1/\lambda) = 1.11$ mm$^{-1}$ is a boundary between the first and second acoustic Brillouin zones (ABZ). Part "1" of the dispersion curve represents AW in the first ABZ, and part "2" represents AW in the second ABZ. The velocities of AW in the 1$^{st}$ and 2$^{nd}$ ABZ may be calculated as $S_1 = 2d \cdot F_1 = 2.943$ km/s and $S_2 = 2d \cdot F_2 = 3.303$ km/s.

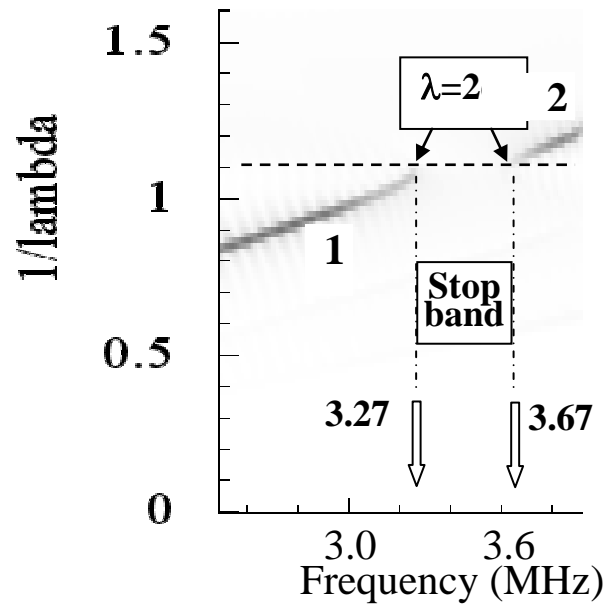

Fig. 2. Dispersion of the zero antisymmetric mode in the ferroelectric inversely poled superlattice fabricated in the ZX-cut 0.5-mm-thick lithium niobate with domain length d = 0.45 mm. The acoustic wave cannot propagate within the stop-band in the range of 3.27 to 3.67 MHz, where the dispersion curve is absent as no solutions are possible due to destructive diffraction from the domain walls.

The rf-current flowing through FPS is calculated by using the same equations (2) and (3). For the known crystallographic orientations of the plate and AW propagation direction, such as in fig. 1, the material constants may be written as certain quantities: $c^E_{ijkl} = c$, $e^{-+}_{ikl}(x) = e^{-+}(x)$ and $\varepsilon^S_{ij} = \varepsilon_0 \varepsilon$, where $\varepsilon_0$ is the vacuum permittivity. Solving equations (2), (3) for the electric field $E$ along with the boundary conditions of zero $E$ at the crystal metalized ends, $E(x = -+Nd) = 0$, one finds a connection between the $E$ and dielectric displacement field $D$.

$$E(x, \omega) = \frac{D(\omega)}{\varepsilon_0 \varepsilon}[1 - \frac{e^{-+}(x)}{|e|} \cdot \frac{Sin(k \cdot x)}{Sin(k \cdot N \cdot d / \sqrt{(1+K^2)})}] \quad , \quad (4)$$

where $K^2 = (e^2 / c\varepsilon_0 \varepsilon)$ is squared piezoelectric coupling coefficient, $|e|$ is positive piezoelectric constant, $k = (\omega/S)$ is wave number. The integration of the $(-E(x, \omega))$ specified by equation (4) over the superlattice length gives the applied voltage $V(\omega) = V_0 \exp(i\omega t)$, which allows for finding a dielectric displacement field $D(\omega)$ as a function of $V(\omega)$. Then a density of rf current $J(\omega,k,N,K)$ flowing through IPS is calculated from the known $D(\omega)$ as a total displacement current:

$$J = -\frac{i\omega\varepsilon_0\varepsilon \cdot V(\omega)}{2Nd}[1 - \frac{2(1+K^2)}{kNd} \cdot \frac{Sin(kd/2)}{Sin(Nkd/\sqrt{(1+K^2)})} \cdot \sum_{m=1}^{N}(-1)^m Sin\frac{(2m-1)kd}{2}]^{-1} , \quad (5)$$

where the first term gives a pure displacement current and the second term represents a contribution of the multidomain FPS. The second term in equation (5) is valid for a propagating AW. Consequently, the fact of a stop-band at the ABZ boundary leads to corresponding splitting in the second term of equation (5), which must be computed for the two modes separately in the 1$^{st}$ and 2$^{nd}$ ABZ, respectively. In a real crystal, AW attenuation is taken into account by a complex wave number $k = (k' - i \cdot k'')$ and corresponding quality factor $Q = (k' / 2 \cdot k'')$. The dependency of $J$ on frequency computed by equation (5) is presented in fig. 3.

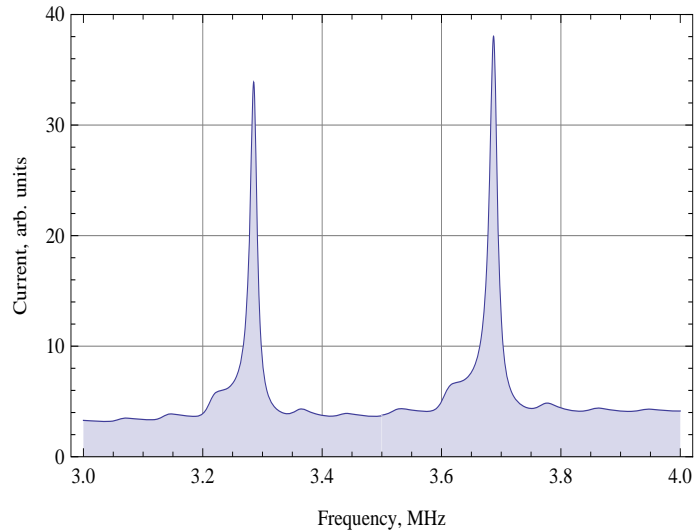

Fig. 3. Computer simulation of rf-current versus frequency from the multidomain superlattice in the ZX-cut LiNbO$_3$ wafer with 44 domain pairs of 0.45-mm-long each. $K^2 = 0.11$, $Q = 105$.

The parameters of the experimental specimen were used for that computation: N = 44; d = 0.45 mm; Q = 105; and $K^2 = 0.11$ for propagating acoustic mode[19].

**3.** *Experiment and discussion.* The sample LN-ZX-MD3, 1 in fig.1, was fabricated out of an optical grade single crystal 0.5-mm-thick LiNbO$_3$ wafer by MTI Corporation with two surfaces polished; the standard Z-cut orientation with orientation tolerance ±0.5° or better, surface quality <5A (by AFM). The pattern of inversely poled domains was fabricated by applying an external constant voltage to the sample with the 0.45-mm period. The resultant superlattice has N = 44 pairs of domains, and is 39.6-mm long along the X-axis, sample width is 29.5 mm along the Y-axis. The micro-picture of the sample taken through polarized microscope revealed a good pattern of the inversely poled domains with the uniform thickness of 0.45 mm of all ferroelectric domains. The experimental measurements are taken at room temperature from the sample sitting in a metal grounded housing to minimize the influence of air rf-fields on the data collected.

The bursts of input voltage $V(\omega)$ are generated by the Agilent Function Generator 33250A. The output rf-current $J$ is collected by the electrode 3 and further is analyzed by the digital storage oscilloscope Tektronix TDS-2024B, and computer. The spectra of applied voltage and output current are presented in Fig. 4, plots (1) and (2), respectively. The current has two

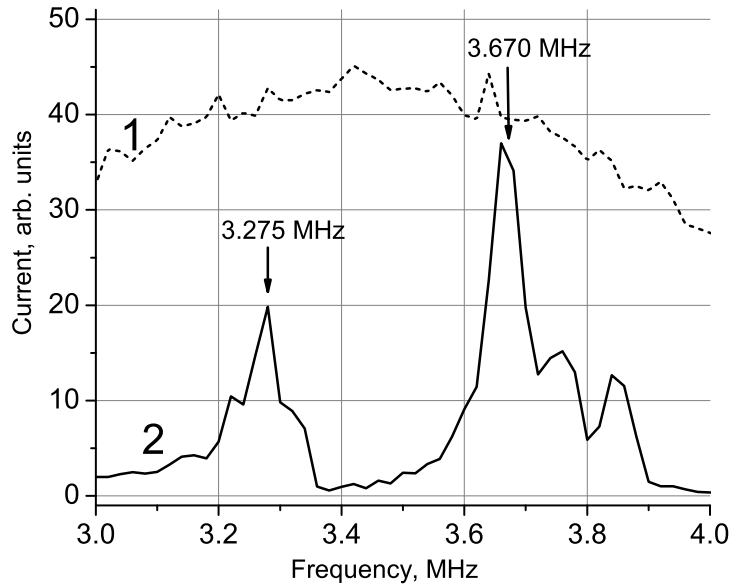

Fig. 4. Spectra of 3-cycle-burst at input electrode 2 of fig.1 (plot 1) and output rf-current detected by electrode 3 of fig.1 (plot 2) from the sample LN-ZX- MD3-AEC.

Two maxima at 3.275 MHz in the 1$^{st}$ ABZ and 3.670 MHz in the 2$^{nd}$ ABZ are indicated by the arrows, the stop band falls between the arrows.

strong peaks at $F_1$ = 3.275 MHz in 1$^{st}$ ABZ and $F_2$ = 3.67 MHz at 2$^{nd}$ ABZ. The experimental plot (2) in fig. 4 is in agreement with theory of fig.3; and both, theory and experiment of figs.3,4 are in good agreement with the calculated stop-band frequencies in fig.2. The respectively small current ripples in figures 3 and 4 are due to diffraction from all domains.

4. ***In conclusion***: The two acousto-electric resonances at ABZ boundary are basic physical property of a piezoelectrically active FPS. It may strongly influence a performance of various applications including multidomain ultrasonic transducers for medicine and nondestructive testing, oscillators and filters for mobile communication, and nano- and micro-electronic elements with polarization reversal. The computations show that even a few domains will demonstrate two acousto-electric resonances. Consequently, the electro-acoustic and inverse acousto-electric transformations may strongly depend on a spectrum of an applied electric voltage. The maximum efficiency of a transformation from electro-magnetic into acoustic energy and vice versa will be at the acousto-electric resonance frequencies $F_1$ and $F_2$. The various experimental details such as metalized surface of a crystal and matching circuits for increasing acousto-electric transduction may produce a frequency-dependent distortion to the two-frequency resonances, which in turn would decrease practical efficiency of the applications. That is why, it is important to design the applications operating in concert with the two native resonances of a given superlattice.

**Acknowledgement**: This work is made possible in part due to the research grant "Nonlinear vibrations of piezoelectric resonators", UM, 2011.